\documentclass[10pt]{article}
\usepackage{graphicx,amssymb,amsmath,amsfonts}

\newcommand{\tr}{{\rm tr}} 
\newcommand{\imag}{i}
\newcommand{\ve}[1]{\boldsymbol{#1}} % vector in normal size

\newcommand{\Fne}{F^{\boldsymbol{n}, \epsilon}}
\newcommand{\Fme}{F^{\boldsymbol{m}, \epsilon}}

\newcommand{\wne}{w_{\ve{n}, \epsilon}}
	
\newcommand{\C}[1]{C(\ve{#1},\epsilon)}	
\newcommand{\id}{{\rm 1 \mkern-4.5mu \nonscript\mkern-.2mu I}}

\newcommand{\absatz}{\vspace{.1in} \noindent}

\newtheorem{definition}{Definition}
\newtheorem{proposition}{Proposition}
\newtheorem{theorem}{Theorem}
\begin{document}
\begin{flushright}
{\footnotesize  pp. 195-203 in T.Placek and J. Butterfield (eds.): 
{\em Non-locality and Modality}, 
\\ NATO Science Series, Kluwer Academic Publishers, Dordrecht, 2002.
}
\end{flushright}

\begin{center}
\vspace{.3in}

\noindent{\Large\bf 
A Kochen-Specker Theorem \\ for Unsharp Spin 1 Observables}

\vspace{.2in}
{\sc Thomas Breuer\footnote{Work partially supported by the Austrian Fonds zur
F\"{o}rderderung der wissenschaftlichen Forschung, Spezialforschungsbereich SFB F012 at
the Universit\"{a}t Salzburg.}
} \\

\vspace{.2in} 
Department of Computer Science, FH Vorarlberg\\
A-6850 Dornbirn, Austria\\
thomas.breuer@fh-vorarlberg.ac.at
\end{center}

\vspace{.3in}
\begin{abstract}
Unsharp spin observables are shown to arise from the fact that a residual 
uncertainty about the actual alignment of the measurement device remains. 
If the uncertainty is below a certain level, and if the distribution 
misalignments is covariant under rotations, a Kochen-Specker theorem
for the unsharp spin observables follows: There are finite sets of directions 
such that not all the unsharp spin observables in these directions can consistently 
be assigned approximate truth-values in a non-contextual way.  
\end{abstract}

\vspace{.2in}
\section{Introduction}
The Kochen-Specker (KS) theorem establishes that not all measurement outcomes predicted
by quantum mechanics can result from detecting hypothetically predetermined values of the
observables. Recently, there have been debates whether or not this result is relevant for 
finite-precision measurements \cite{Pitovsky 1985, Meyer 1999, Kent 1999, 
Clifton and Kent 2000, 
Cabello et al 1997, Cabello 2001, Simon et al 2000, 
Simon et al 2001, Mermin 1999, Havlicek et al 2000, Appleby 2001}. The goal of this paper 
is (1) to give a sound description of measurements with a residual uncertainty about the 
actual orientation of the measurement device, (2) to establish a KS-type
no-hidden-variables theorems for this kind of finite precision measurements. For the sake of
definiteness I will discuss these issues for a spin 1 particle, but similar results hold 
for arbitrary observables in a Hilbert space of dimension 3 or higher.   
 
In Section~\ref{sec-Finite Precision Measurements of Spin 1 Observables} 
I will construct POV-observables representing finite precision 
measurements of a spin 1 particle. I will also discuss some important properties of 
these unsharp spin observables. In 
Section~\ref{sec-The KS-theorem for Unsharp Spin 1 Observables} I will prove a KS-theorem for these 
unsharp spin observables. The paper concludes with a discussion of the relation of these 
results to the current debate on the validity of KS-theorems for finite precision 
measurements.  

\section{\label{sec-Finite Precision Measurements of Spin 1 Observables}
Finite Precision Measurements of Spin 1-Observables}
The sharp spin 1 observables in $\ve{x}, \ve{y}$, and $\ve{z}$  
direction are given by the three 
three-dimensional Pauli matrices $S_{\ve{x}}, S_{\ve{y}}, S_{\ve{z}}$, 
each of which has eigenvalues 1, 0, and -1. 
For example $S_{\ve{z}}$ is given by 
\begin{equation}
S_{\ve{z}} = 
\left(
\begin{array}{ccc}
1 & 0 & 0 \\
0 & 0 & 0 \\
0 & 0 & -1
\end{array}
\right). 
\end{equation}
Denote the eigenvectors of the spin matrix $S_{\ve{z}}$ by 
$\psi_{\ve{z},1},\psi_{\ve{z},0},\psi_{\ve{z},-1}$ and 
the  corresponding eigenprojectors by   
$P_{\ve{z},1}:=\vert\psi_{\ve{z},1}\rangle\langle\psi_{\ve{z},1}\vert$ 
and similarly for $P_{\ve{z},0}$ and 
$P_{\ve{z},-1}$. 
The $P_{\ve{z},i}$ are sharp spin properties. For example, 
\begin{equation}
\label{eq-Pz}
P_{\ve{z},1} = 
\left(
\begin{array}{ccc}
1 & 0 & 0 \\
0 & 0 & 0 \\
0 & 0 & 0
\end{array}
\right). 
\end{equation}
Similar notation will be used for the $\ve{x}$- and the $\ve{y}$-axes. 

For an arbitrary direction $\ve{n}$ the sharp spin 1 observable is 
$S_{\ve{n}}:=\ve{n}\cdot \ve{S}$, where $\ve{S}$ is the Pauli vector $(S_{\ve{x}}, 
S_{\ve{y}}, S_{\ve{z}})$.
$S_{\ve{n}}$ also has eigenvalues 1, 0, and -1. Let $\psi_{\ve{n},i}$ and $P_{\ve{n} ,i}$ be 
the eigenstates and eigenprojectors of $S_{\ve{n}}$ corresponding to the eigenvalues 
$i=1,0,-1$.  
The sharp spin observable $S_{\ve{n}}$ in direction $\ve{n}$ can be represented as a 
projection valued (PV)-measure on the value space $\Omega=\{1,0,-1\}$, which associates
to each element $i$ of the value space $\Omega$ the projector $P_{\ve{n}, i}$.

Now assume we are not sure that we actually measure the spin in the direction $\ve{n}$
we want. We only know that the directions $\ve{m}$ of actual spin measurements is distributed with 
a density $\wne(\ve{m})$ around the intended direction $\ve{n}$. The probability that 
such an imprecisely specified measurement yields 
an outcome +1 when the spin 1 system is prepared in some pure state $\psi$ is 
$\mbox{\rm Prob}_\psi^{\ve{n},\epsilon}(+1) = \int_{S^2}d\Omega (\ve{m}) \wne(\ve{m}) \tr (P_\psi P_{\ve{m} , +1})
= \tr ( P_\psi \int_{S^2}d\Omega (\ve{m}) \wne(\ve{m})  P_{\ve{m} , +1} )$,
where $d\Omega$ is the Lesbesgue-measure of the sphere.   
Defining 
\begin{eqnarray}
\label{def-Fne}
F^{\ve{n},\epsilon}(i):=\int_{S^2} w_{\ve{n},\epsilon}(\ve{m}) P_{\ve{m},i} \, d\Omega(\ve{m}).
\end{eqnarray}
the probability of outcome $i$ can be written as 
$$\mbox{\rm Prob}_\psi^{\ve{n}, \epsilon}(i) = \tr \left(P_\psi F^{\ve{n},\, \epsilon}(i)\right).$$ 

From (\ref{def-Fne}) it is obvious that the $\Fne$ are positive self-adjoint operators 
satisfying $0 \leq \Fne \leq \id$. But they are not projectors since 
$\Fne(i)\not= \Fne(i)^2$. 
The $\Fne(i)$ form a resolution of the identity,
\begin{equation}
\label{eq-resolution of identity}
\Fne(1) + \Fne(0) + \Fne(-1) = \id,
\end{equation}
which follows from (\ref{def-Fne}) and 
$P_{\ve{n},1} + P_{\ve{n},0} + P_{\ve{n},-1}= \id$.
Thus we have a positive operator valued measure (POVM) $\Fne$ 
which associates
to each element $i$ of the value space $\Omega$ the projector $\Fne(i)$.
Because of (\ref{eq-resolution of identity})
this measure is normalised. POV-measures are the standard
tool for describing realistic experiments \cite{Busch et al 1995, Holevo 1982}.

\begin{proposition}
If the distributions $\wne$ of apparatus misalignments transform covariantly under rotations, 
$\wne(R\ve{m})=w_{R^{-1}\ve{n}, \epsilon} (\ve{m})$, then the unsharp 
spin properties $\Fne$ transform covariantly under rotations,
\begin{eqnarray}
\label{eq-covariance of Fne}
D^1(R) \, F^{\ve{n},\, \epsilon}(i) \, D^1(R)^{-1} = F^{R^{-1}\ve{n}, \epsilon}(i),
\end{eqnarray}
where $D^1$ is the spin 1-representation of the rotation group.
\end{proposition}
To see this note that the sharp spin properties $P_{\ve{m} \, i}$ are angular momentum 
operators and therefore transform covariantly \cite[p. 70]{Busch et al 1995}, 
\begin{eqnarray}
\label{eq-covariance of Pn}
D^1(R) \, P_{\ve{m}, i} \, D^1(R)^{-1} = P_{R^{-1}\ve{m}, i}.
\end{eqnarray}
Using this and and the definition of the unsharp spin properties (\ref{def-Fne}), 
we obtain 
$D^1(R) \, \Fne(i) \, D^1(R)^{-1} = 
\int_{S^2} \wne(\ve{m})P_{R^{-1}\ve{m} , i}d \ \Omega(\ve{m})$.  
Introducing new coordinates $\ve{m}'=R^{-1}\ve{m}$, and using the rotation invariance of 
$d \, \Omega$ and $\wne(R\ve{m})=w_{R^{-1}\ve{n}, \epsilon} (\ve{m})$, the right hand
side is seen to equal $F^{R^{-1}\ve{n}, \epsilon}(i)$, 
which establishes (\ref{eq-covariance of Fne}).

Since the $\Fne(i)$ transform covariantly under rotations, they are angular momentum properties
and can be regarded as spin properties with the same justification as the sharp spin properties $P_{\ve{n},i}$.
This is in line with Weyl's idea of defining observables by their transformation properties under some kinematic 
group.

\begin{proposition}
\label{prop2}
If the distribution $\wne$ of measurement errors transforms covariantly under rotations, 
$\wne(R\ve{m})=w_{R^{-1}\ve{n}, \epsilon} (\ve{m})$, then every eigenvector
of the sharp spin properties $P_{\ve{m} \, i}$ is also an eigenvector of the unsharp 
spin properties $\Fne(i)$. Since the sharp spin properties $\{P_{\ve{n},i}\}_{\, i=1,0,-1}$
have simultaneous eigenvectors and commute, this is also the case for the unsharp spin 
properities $\{\Fne(i)\}_{\, i=1,0,-1}$. 
\end{proposition}

To see this, we proceed in two steps. In a first step we consider the special case that the
intended measurement direction $\ve{n}$ is the $\ve{z}$-direction.  
The eigenvectors of $P_{\ve{z},i}$ are the vectors in the $\ve{x}, \ve{y},$ or $\ve{z}$-directions. We will show
that $\Fne$ has the same eigenvectors. 

Using the explicit form of the spin matrices 
$S_{\ve{x}}, S_{\ve{y}}, S_{\ve{z}}$, one can calculate the $P_{\ve{m} , i}$. For $i=1$ and $\ve{m}$ 
in polar coordinates $(\theta, \phi)$ it turns out that $P_{\ve{m}(\theta, \phi),1}$ equals 
{\small
$$
\left(
\begin{array}{ccc}
 \cos (\frac{\theta}{2})^4, & 
\frac{1}{2 \sqrt{2}} \, e^{- \imag \, \phi} \, (1 + \cos (\theta)) \, \sin (\theta) & 
\frac{1}{4} \, e^{2\imag \phi} \sin (\theta)^2\\
\frac{1}{2\sqrt{2}} \, {e^{\imag \, \phi}\, (1 + \cos (\theta)) \, \sin (\theta)} &
\frac{1}{2} \, \sin (\theta)^2& 
\sqrt{2} e^{-\imag \phi} \, \cos (\frac{\theta}{2})\, \sin (\frac{\theta}{2})^3 \\ 
\frac{1}{4} \, e^{-2\imag \phi} \sin (\theta)^2&
\sqrt{2} e^{\imag \phi} \, \cos (\frac{\theta}{2})\, \sin (\frac{\theta}{2})^3 & 
\sin (\frac{\theta}{2})^4
\end{array}
\right) 
$$
 }

The covariance property of $\wne$ implies that $\wne$ is invariant under rotations around 
$\ve{n}$: If $R$ is a rotation around $\ve{n}$, then 
$\wne(R\ve{m})=w_{R^{-1}\ve{n}, \epsilon} (\ve{m})=\wne(\ve{m})$.
Since
$w_{\ve{z}, \epsilon}$ is invariant under all rotations around the $\ve{z}$-axis, we can write
$w_{\ve{z}, \epsilon}(\ve{m}) = w_{\ve{z}, \epsilon}(\ve{m}(\theta,\phi)) =: w_{\ve{z}, \epsilon}(\theta)$, 
where $(\theta,\phi)$ are the polar coordinates of the unit vector in
the direction $\ve{m}$. Thus we have  
$$\int_0^{2 \pi} w_{\ve{z}, \epsilon}(\theta,\phi) \, P_{\ve{m}(\theta, \phi), 1} \, d \phi = 
w_{\ve{z}, \epsilon}(\theta) \int_0^{2 \pi}  P_{\ve{m}(\theta, \phi), 1} \, d \phi.$$
Evaluating the integral on the right hand side yields the matrix
{\small
$$
\left(
\begin{array}{ccc}
 2\pi  \cos (\frac{\theta}{2})^4, & 0 & 0 \\
0 &
\pi \, \sin (\theta)^2 & 0 \\
0 & 0 & 2\pi \, \sin (\frac{\theta}{2})^4
\end{array}
\right). 
$$
 }

Therefore
{\small
\begin{eqnarray}
F^{\ve{z}, \epsilon}(1) & = & 
\int_{S^2} w_{\ve{z}, \epsilon}(\ve{m}) P_{\ve{m}, i} \, d\Omega(m) \nonumber \\
& = & \int_0^\pi d\theta \sin (\theta) 
\int_0^{2 \pi}  d \phi \, w_{\ve{z}, \epsilon}(\theta,\phi) \, P_{\ve{m}(\theta, \phi), 1} \nonumber \\
& = & 
\left(
\begin{array}{ccc}
 \alpha_1 & 0 & 0 \\
0 &
\alpha_2 & 0 \\
0 & 0 & \alpha_3
\end{array}
\right). \label{eq-Fzeplus}
\end{eqnarray} 
} 
Similarly one can derive 
\begin{equation}
\label{eq-Fze}
F^{\ve{z}, \epsilon}(0) = 
\left(
\begin{array}{ccc}
\alpha_2 & 0 & 0 \\
0 &
\alpha_4 & 0 \\
0 & 0 & \alpha_2
\end{array}
\right) \qquad
F^{\ve{z}, \epsilon}(-1) = 
\left(
\begin{array}{ccc}
\alpha_3 & 0 & 0 \\
0 &
\alpha_2 & 0 \\
0 & 0 & \alpha_1
\end{array}
\right), 
\end{equation}
where 
\begin{eqnarray}
\alpha_1 & = & 2\pi \int_0^\pi d\theta 
\, w_{\ve{z}, \epsilon}(\theta) \sin (\theta)  \cos (\theta/2)^4 \nonumber \\
\alpha_2 & = & \pi \,\int_0^\pi d\theta  
\, w_{\ve{z}, \epsilon}(\theta) \sin (\theta) \sin (\theta)^2 \nonumber \\
\alpha_3 & = & 2\pi \, \int_0^\pi d\theta  
\, w_{\ve{z}, \epsilon}(\theta) \sin (\theta) \sin (\theta/2)^4. \label{eq-eigenvalues of Fne} \\
\alpha_4 & = & 2\pi \, \int_0^\pi d\theta  
\, w_{\ve{z}, \epsilon}(\theta) \sin (\theta) \cos (\theta)^2. \nonumber
\end{eqnarray}
The  $F^{\ve{z}, \epsilon}(i)$ are diagonal matrices. Therefore the unit vectors in the 
$\ve{x}, \ve{y},$ and $\ve{z}$-directions 
are eigenvectors of all $F^{\ve{z}, \epsilon}(i)$. This finishes the proof of
Proposition~\ref{prop2} for the special case $\ve{n}=\ve{z}$.

In the second step we consider the general case that we want to measure spin
in an arbitrary direction $\ve{n}$. First we will show that the eigenvectors of
$\Fne(i)$ are $D^1(R)^{-1}\ve{x}, D^1(R)^{-1}\ve{y}, D^1(R)^{-1}\ve{z}$, where $R$ is a 
rotation fulfilling $R^{-1}\ve{n}=\ve{z}$. Using the covariance property 
(\ref{eq-covariance of Fne}) and the fact that $\ve{x}, \ve{y}, \ve{z}$ are eigenvectors 
of the $F^{\ve{z}, \epsilon}(i)$ we verify 
$\Fne(i) D^1(R)^{-1}\ve{x} = 
D^1(R)^{-1} F^{\ve{z}, \epsilon}(i) D^1(R) D^1(R)^{-1}\ve{x} = $ 
$D^1(R)^{-1} F^{\ve{z}, \epsilon}(i) \ve{x} = $
$D^1(R)^{-1} \alpha \ve{x} = $
$\alpha D^1(R)^{-1} \ve{x}$, which implies 
that $D^1(R)^{-1}\ve{x}$ is an eigenvector of $\Fne(i)$.
Similarly we can check that $D^1(R)^{-1}\ve{y}$ and $D^1(R)^{-1}\ve{z}$ 
are eigenvectors of $\Fne(i)$.
Since $\Fne(i)$ only has three eigenvectors,  $D^1(R)^{-1}\ve{x},$ 
$D^1(R)^{-1}\ve{y}, D^1(R)^{-1}\ve{z}$ are the only eigenvectors of $\Fne(i)$.

Next we will show that $D^1(R)^{-1}\ve{x}, D^1(R)^{-1}\ve{y}, D^1(R)^{-1}\ve{z}$ are the 
eigenvectors of $P_{\ve{n},i}$.
Using the covariance property (\ref{eq-covariance of Pn}) of the sharp spin observables 
and the fact that $\ve{x}, \ve{y}, \ve{z}$ are eigenvectors 
of the $P_{\ve{z},i}$ with eigenvalues $\lambda=1,0,-1$ we verify 
$P_{\ve{n},i} D^1(R)^{-1}\ve{x} = 
D^1(R)^{-1} P_{\ve{n},i} D^1(R) D^1(R)^{-1}\ve{x} = $ 
$D^1(R)^{-1} P_{\ve{n},i} \ve{x} = $
$D^1(R)^{-1} \lambda \ve{x} = $
$\lambda D^1(R)^{-1} \ve{x}$, which implies 
that $D^1(R)^{-1}\ve{x}$ is an eigenvector of $P_{\ve{n},i}$.
Similarly we can check that $D^1(R)^{-1}\ve{y}$ and $D^1(R)^{-1}\ve{z}$ 
are eigenvectors of $P_{\ve{n},i}$.
Since $P_{\ve{n},i}$ only has three eigenvectors,  $D^1(R)^{-1}\ve{x},$ 
$D^1(R)^{-1}\ve{y}, D^1(R)^{-1}\ve{z}$ are the only eigenvectors of $P_{\ve{n},i}$.
Thus the $P_{\ve{n},i}$ and the $\Fne(i)$ all have the same three eigenvectors.
This establishes Proposition~\ref{prop2} for the general case.

\begin{proposition}
The eigenvalues of the unsharp spin properties $\Fne(i)$ are in the set $\{\alpha_1, \ldots, \alpha_4 \}$, 
where the $\alpha_i$ are given by equations (\ref{eq-eigenvalues of Fne}).
The eigenvalues of each $\Fne(i)$ add up to $1$.
\end{proposition}
By the covariance property (\ref{eq-covariance of Fne}) of the $\Fne(i)$, 
the $\Fne(i)$ have the same eigenvalues as 
$F^{\ve{z}, \epsilon}(i)$. The eigenvalues of the $F^{\ve{z}, \epsilon}(i)$ are the $\alpha_i$  
of equations (\ref{eq-eigenvalues of Fne}). Taking into account that
$1=\int_{S^2} d\Omega (\ve{m}) \wne(\ve{m})=
2\pi \int_0^\pi d\theta \sin(\theta) \wne(\theta)$ one easily verifies from  
(\ref{eq-eigenvalues of Fne}) that $0 \leq \alpha_i \leq 1$ and 
$\alpha_1 + \alpha_2 + \alpha_3 = 1$, $\alpha_2 + \alpha_4 + \alpha_2 = 1$.

\absatz 
To give an explicit example, assume that the
spin directions actually measured are uniformly distributed over $\C{n}$, 
the set of directions
deviating from $\ve{n}$ by less than an angle $\epsilon$. Denote by $A$ the size of
$\C{n}$ on the unit sphere. So $\wne$ is  $1/A$ times the characteristic function of 
$\C{n}$. The $F^{\ve{z}, \epsilon(i)}$ have the form (\ref{eq-Fzeplus}) and (\ref{eq-Fze})
with
\begin{eqnarray}
\alpha_1 & = & \frac{1}{24} \left(15 + 8 \cos(\epsilon) + \cos(2\epsilon) \right) \nonumber \\
\alpha_2 & = & \frac{1}{3} \left(2 + \cos(\epsilon)\right)  \sin(\epsilon/2)^2 \nonumber \\
\alpha_3 & = & \frac{1}{3} \sin(\epsilon/2)^4 
\label{eq-eigenvalues of Fne for uniform distribution} \\
\alpha_4 & = & \frac{1}{6} \left(3 + 2 \cos(\epsilon) + \cos(2\epsilon)\right) . \nonumber
\end{eqnarray}
Observe that, as the measurement inaccuracy $\epsilon$ tends to zero, two eigenvalues 
($\alpha_2$ resp. $\alpha_3$) of each $F^{\ve{z},\epsilon}(i)$ go to zero, 
one eigenvalue ($\alpha_1$ resp. $\alpha_4$) goes to one. Comparing (\ref{eq-Fzeplus}) to (\ref{eq-Pz}) 
we see that the unsharp spin properties
$F^{\ve{z},\epsilon}(i)$ converge to the sharp spin properties $P_{\ve{z},i}$ as the measurement 
inaccuracy goes to zero.

Additionally, given 
any vector which is an eigenvector of the $\Fne(i)$, for one of the $\Fne(i)$ this vector 
has an eigenvalue close to one, and for the two other $\Fne(i)$, this vector has an 
eigenvalue close to zero. For example, the vector $(0,1,0)$ is an eigenvector of 
$F^{\ve{z},\epsilon}(0)$ with eigenvalue $\alpha_4$ (which is close to one), 
and it is an eigenvector of $F^{\ve{z},\epsilon}(1)$ and $F^{\ve{z},\epsilon}(-1)$ 
with eigenvalue $\alpha_2$ (which is close to zero).

\section{\label{sec-The KS-theorem for Unsharp Spin 1 Observables}
The KS-theorem for Unsharp Spin 1 Observables}
Determining the result $i\in\{1,0,-1\}$ of a {\em sharp} spin measurement in direction 
$\ve{n}$ is picking one of the sharp spin properties $\{P_{\ve{n},i}\}_{i=1,0,-1}$ and 
assigning it the truth value 1. 
Since the sharp spin properties are projectors 
$P_{\ve{n},i}:=\vert\psi_{\ve{n},i}\rangle\langle\psi_{\ve{n},i}\vert$
they can be identified with the rays $\psi_{\ve{n},i}$. So, assigning the value 1 to one of 
the $P_{\ve{n},i}$ and the value 0 to the other two, is equivalent to assigning the colour
T (true) to one of the rays $\psi_{\ve{n},i}$ and the colour F (false) 
to the two other rays. 
The traditional KS-proofs show that for certain sets of directions
this colouring rule cannot be satisfied.

Determining the result $i\in\{1,0,-1\}$ of an {\em unsharp} spin measurement in 
direction $\ve{n}$ is picking one of the unsharp spin properties 
$\{\Fne(i)\}_{i=1,0,-1}$ and assigning it the truth value 1. 
But the unsharp spin properties are not projectors and therefore cannot readily be
identified with rays. To arrive at a colouring rule for rays we have to proceed in 
a different way. 

Assume for example that one intends to measure spin in 
direction $\ve{z}$ and
result 0 occurs. This means that the unsharp spin property $F^{\ve{z},\epsilon}(0)$ 
is realised, whereas $F^{\ve{z},\epsilon}(1)$ and $F^{\ve{z},\epsilon}(-1)$ are not 
realised. Accordingly we assign the colour AT (almost true) to the ray $(0,1,0)$, 
which is the eigenvector of $F^{\ve{z},\epsilon}(0)$ with eigenvalue close to 1. 
To $(1,0,0)$ and $(0,0,1)$ we assign the colour AF (almost false) because they are eigenvectors 
of $F^{\ve{z},\epsilon}(0)$ with eigenvalue close to 0. 
Had the outcome been 1, we would have assigned AT to $(1,0,0)$ and AF to $(0,1,0)$ 
and to $(0,0,1)$. This is achieved by the following value assignment. 
\begin{definition}
\label{def-colouring}
Fix some unsharpness tolerance $0.5>\delta \geq 0$.
If the outcome of a spin measurement with some intended direction $\ve{n}$ is 
some $i\in\{1,0,-1\}$, 
then the ray of the eigenvector of $\Fne(i)$ corresponding to an eigenvalue larger or 
equal to $1-\delta$ gets colour AT, and the rays of eigenvectors corresponding to 
eigenvalues smaller or equal to $\delta$ get colour AF. Rays corresponding
to some eigenvalue between $\delta$ and $1-\delta$ are assigned a third colour or no colour.
\end{definition} 
$\delta>0$ is an unsharpness tolerance below which an eigenvalue counts as ``almost zero".
An eigenvalue above $1-\delta$ counts as ``almost one". The exact level is a matter of taste,
and our results do not depend on the exact level. But certainly $\delta$ should be smaller 
than 0.5, since otherwise some values would simultaneously be counted as almost zero
and almost one. 

In a way this assignment of approximate truth values is conservative: 
a state assigns approximate truth values only to those unsharp spin properties 
of which it is an eigenstate. 
Note that if $\Fne(i)$ possesses an approximate truth value, it does possess it 
without dispersion. 
Contrast this with a situation
where the spin in some direction $\ve{n}'$ close to $\ve{n}$ has value $i$; the spin value 
in direction $\ve{n}$ is not dispersion-free.  

Note that the intended measurement direction $\ve{n}$ need not be among 
the coloured rays. 
An example: If $\ve{n}=\ve{x}$, the rays which are coloured are the 
normalised eigenvectors of the $F^{\ve{x},\epsilon}(i)$. By Proposition~\ref{prop2} these
are exactly the rays of the eigenvectors of $P_{\ve{x},i}$. The eigenvectors of the 
$P_{\ve{x},i}$ are by definition the eigenvectors of $S_{\ve{x}}$, 
namely $(1,-\sqrt{2},1)/2,$ $(-1,0,1)/\sqrt{2}$, and 
$(1,\sqrt{2},1)/2$. The direction $\ve{x}$ is not coloured. 
Note also that we are colouring complex rays, not real rays.

In Proposition~\ref{prop2} we have seen that for a fixed intended measurement direction
$\ve{n}$ the $\{\Fne(i)\}_{i=1,0,-1}$ have the same eigenvectors. 
If the measurement inaccuracy is sufficiently small---or to me more precise: 
if the density $\wne(\ve{m})$ of apparatus misalignments has enough probability mass 
sufficiently close to the intended measurement direction $\ve{n}$---then 
the eigenvalues $\alpha_1$ and
$\alpha_4$ of equations (\ref{eq-eigenvalues of Fne}) will be larger than $1-\delta$, 
whereas $\alpha_2$ and $\alpha_3$ will be smaller than $\delta$. If this is the case,
for each direction $\ve{n}$,  
exactly one ray in the orthogonal triad of eigenvectors of the $\Fne(i)$ will
get colour AT, and two rays will get colour AF. 

For example if we assume apparatus misalignments to be uniformly distributed over 
the set of directions deviating by less than an angle $\epsilon$ and choose $\delta=0.1$, 
then we can calculate from equations (\ref{eq-eigenvalues of Fne for uniform distribution}) 
that for $\epsilon$ smaller than $0.459=26.3^{\rm o
}$ condition (2) is fulfilled and 
$\alpha_1,\alpha_4$ will be larger than $0.9$,  
$\alpha_2, \alpha_3$ will be smaller than $0.1$. So for $\delta=0.1$, 
if the measurement inaccuracy $\epsilon$ is smaller than 0.459, 
then in all orthogonal tripods one ray will be coloured AT and two rays will be 
coloured AF.       

Let $N$ 
denote a set of directions, and $F(N,\epsilon)$ the set of positive operators in the ranges 
of the $\Fne$ with $n\in N$. Let $V(N,\epsilon)$ the set of rays of the eigenvectors
of the $\Fne$ with $n\in N$. Then, a hidden variable model can uniquely determine 
the measured values of spin measurements in all directions of $N$ only if there exists
a colouring $V(N,\epsilon) \rightarrow \{AT,AF\}$ which associates for all 
directions $\ve{n}\in N$ the colour AT to one of the eigenvectors of the $\Fne(i)$ and 
the colour AF to the two other eigenvectors.

One ray can be eigenvector of spin properties $\Fne(i), \Fme(i)$ 
in different directions $\ve{n}, \ve{m}$. Non-contextuality of the hidden variable-model 
implies that such a ray is assigned a unique colour.
Now the KS-theorem for unsharp spin observables follows in exactly the same way as the one
for sharp observables. 
In every orthogonal tripod one of the rays is constrained to
get the colour AT, the other two rays get AF. For the KS-sets of tripods such a colouring 
is impossible \cite{Kochen and Specker 1967, Peres 1995, Zimba and Penrose 1993}.  
Thus we arrive at the following

\begin{theorem}
\label{theorem 1}
For any unsharpness tolerance $0.5>\delta\geq 0$, if in an unsharp spin 1 measurement \\
(1) the densities of apparatus misalignments transform  covariantly, 
$\wne(R\ve{m})=w_{R^{-1}\ve{n}, \epsilon} (\ve{m})$, and \\ 
(2) the measurement inaccuracy described by the densities $\wne(\ve{m})$ of apparatus 
misalignments is so small that in equations 
(\ref{eq-eigenvalues of Fne}) $\alpha_1$ and
$\alpha_4$  are larger or equal to $1-\delta$, 
while $\alpha_2$ and $\alpha_3$ are smaller or equal to $\delta$, \\
then not all the unsharp spin observables in a KS-set of directions can consistently 
be assigned approximate truth-values in a non-contextual way. 
\end{theorem}

\section{Discussion}
Let me briefly relate this result to the debate on the relevance of the 
KS-theorem for inaccurate measurements. This debate was fueled by claims of 
Meyer \cite{Meyer 1999}, Kent \cite{Kent 1999}, and Clifton and 
Kent \cite{Clifton and Kent 2000}, MKC for short, 
that for finite precision measurements the 
KS-theorem is irrelevant. 

Infinite precision is crucial to the  KS-argument in two ways 
\cite{Held 2000}: (1) It is necessary to the argument that the measured components of one 
tripod are exactly orthogonal. (2) In order to exploit non-contextuality 
it is necessary that two measurements intended to pick out the same observable as member of
two different maximal sets pick out exactly the same direction. MKC show that 
non-contextual hidden variable models can be constructed if we relax the assumption of
infinite precision and allow for an arbitrarily small violation of (2). 
In these models it is not exactly the directions in a KS-set that are 
assigned non-contextual values, but points which approximate them 
arbitrarily closely. 
In fact it is possible to assign values to a {\em dense} set of observables. 
(In our spin 1 case this amounts to a colouring of a dense subset of the unit sphere.)

So the hidden variable-theorist is free to adopt the hypothesis that due to some 
apparatus misalignment instead of the intended observable he measures  
another observable, which cannot be distinguished from the intended observable
by a finite precision measurement. The results of these 
finite precision measurements can be explained by a non-contextual hidden variable model. 

In the MKC model the distribution $\wne(\ve{m})$ of misalignments is 
not rotation covariant: Assume that instead of an intended 
direction $\ve{n}$ the apparatus is in fact aligned in a direction $\ve{m}$ with 
rational coordinates, which is very close to $\ve{n}$. Now let $R$ be a rotation by 
$45^{\mbox{o}}$ around an axis orthogonal to $\ve{m}$. If $\wne(\ve{m})$ were rotation
covariant, in an experiment designed to measure spin in direction $R\ve{n}$ we would 
actually measure spin in direction $R\ve{m}$. So the result of a measurement in intended
direction $R\ve{n}$ would be determined by the colour assigned to the point $R\ve{m}$.
But $R\ve{m}$ cannot have rational 
coordinates if $\ve{m}$ has. So $R\ve{m}$ would not be assigned a colour in Meyer's model 
and the result of a measurement in intended direction $R\ve{n}$ would not be predetermined.
The result of a measurement in intended direction $R\ve{n}$ must therefore be determined
by the colour of some point other than $R\ve{m}$.
Thus the distribution $\wne(\ve{m})$ of misalignments in the MKC-models cannot not be 
covariant under all rotations. In Meyer's model condition (1) of Theorem~\ref{theorem 1} 
is not satisfied. 
  
From the more operational point of view taken in this paper, the observer who wants to 
measure spin in a direction $\ve{n}$ will have an experimental procedure for trying to 
do this as exactly as possible. Simon {\em et al.} \cite{Simon et al 2001} refer to this 
procedure by saying he sets the ``control switch" of his apparatus to the position 
$\ve{n}$. The switch position is all the observer knows about. In an operational sense, 
the physical observable measured is entirely determined by the switch position. However,
there will usually be degrees of freedom of the apparatus which the experimenter cannot
control. This results in an apparatus misalignment of which the experimenter is not aware. 
If he were aware of it he would correct it. Not being aware of the 
misalignment he interprets the outcome produced by the misaligned apparatus as result
of an experiment without misalignment. 

Unlike Simon {\em et al} \cite{Simon et al 2001} I do not describe the 
misalignment by associating hidden variables to the apparatus. Rather I describe the
effects of the misalignment by the 
unsharp spin observables $\Fne$. These are constructed so as to 
yield exactly the statistics of outcomes associated to the switch positions. 
Theorem~\ref{theorem 1} shows that the results of measurements with misalignments
distributed in a rotation covariant way cannot be predetermined by a non-contextual 
hidden variable model. 

\end{document}